\documentclass[11pt]{elsarticle}

\pdfoutput=1

\makeatletter
\def\ps@pprintTitle{%
  \let\@oddhead\@empty
  \let\@evenhead\@empty
  \let\@oddfoot\@empty
  \let\@evenfoot\@oddfoot
}
\makeatother

\usepackage{url}
\usepackage{breakurl}
\usepackage[breaklinks,
            colorlinks = true,
            linkcolor = blue,
            urlcolor  = blue,
            citecolor = blue,
            anchorcolor = blue]{hyperref}

\usepackage{lineno}

\usepackage{graphicx}
\usepackage[margin=1.25in]{geometry}
\usepackage[usenames,dvipsnames]{color}
\graphicspath{ {./images/} }

\newcommand\acp{\begin{center}
\rule[-0.2in]{\hsize}{0.01in}\\\rule{\hsize}{0.01in}\\
\vskip 0.1in Submitted to the  Proceedings\\ 
of the African Conference on Fundamental and Applied Physics
    \vskip 0.05in
    {\it Second Edition, ACP2021, March 7--11, 2022 --- Virtual Event}\\
\rule{\hsize}{0.01in}\\\rule[+0.2in]{\hsize}{0.01in} \\
\end{center}}

\usepackage[firstpage=true]{background}
\backgroundsetup{contents={\parbox{6.5in}{\acp}}, scale=1,placement=top,opacity=1,color=black,position={3.25in,1.2in}}

\usepackage{fancyhdr}
\fancypagestyle{plain}{%
  \fancyhf{}%
  \fancyhead[C]{}
  \fancyfoot[C]{\thepage}
}
\usepackage[rightcaption]{sidecap}
\usepackage{wrapfig}

\fancypagestyle{empty}{%
  \fancyhf{}%
  \fancyhead[C]{{\it ACP2021, March 7--11, 2022 --- Virtual Edition}}
  \fancyfoot[C]{\thepage}
}
\begin{document}

\begin{frontmatter}


\title{On making physics relevant to society in general and to scientists in particular: Closing the epistemic gap}

\author[add1]{Jamal Mimouni\corref{cor1}}
\ead{jamalmimouni@umc.edu.dz}

\address[add1]{Depart. of Physics, Univ. of Constantine1 and CERIST, Algeria}
\begin{abstract}
\noindent 
Physics has a bad press: it is seen by a majority of people as a boring discipline ever since their High School days. There is no glamour to it, just toil and pain, and for many who engaged in it, the end sight is often unemployment. Could it be that physicists don't know how to communicate what their discipline entails to? I will be tackling the problematic of making physics relevant to society and to the scientists in general. I will also be dealing with the methodological and educational aspects of teaching and practicing physics, and the need to close the epistemic gap between physics teaching and the physicist's understanding ... By the way, do physicists understand physics? 
\begin{center}

\textit{“…that statement made me furious, so I started studying physics.”} 
Masotoshi Toshiba in his 1997 Nobel Prize address 
\end{center}
\end{abstract}

\begin{keyword}
Physics and society, didactics of physics, epistemic gap 
\end{keyword}

\end{frontmatter}

%



\section{Introduction}
\label{sec:intro}
\noindent
There is an overall feeling of negativity from the general public towards physics seen as an unyielding and boring discipline, all this feed up by an unfruitful encounter with it during the High School years. This strong feeling goes further to encompass its practitioners as I will argue. This is a hard reality we have to face and correct, and making the general public aware of physics’ centrality in science is a vital first step in that direction. One should make people aware not only of its utilitarian aspect through its contributions to the economy, industry and medicine, but it should include its cognitive aspect along with its mind-boggling discoveries and puzzling open questions. Educating people in the methodology of science, in this era of fake news is also of great importance as physics is the archetypal discipline where science in action might be seen in its purest form or at least the simplest one, even as advances may takes meandrous paths.

We will start by arguing that physics is a fundamental discipline, and indeed the most fundamental one, which simply stated is the gateway to understand the material (I was about to say the “physical”…) world. This is not a self-aggrandizing statement uttered by a physicist, but an obvious fact that we physicists don’t often want to make out of displaced modesty. This of course doesn’t mean that physics is the most important discipline or the “noblest” one, whatever it may mean, as the criteria for evaluating the impact of the field is complex and has also to do with the degree to which society and the economy make a beneficial use of it. In fact for this later title, it looks pretty straightforward that computer science and electronics might be the top contenders\footnote{\label{myfootnote}Both based on physics: Quantum mechanical properties of semiconductors for the computer chips, Maxwell’s equations in a given regime for the other.}.

\section{Physics as Magic}
\label{sec:grate}
\begin{figure}[h]
\centering
\includegraphics[width=8cm]{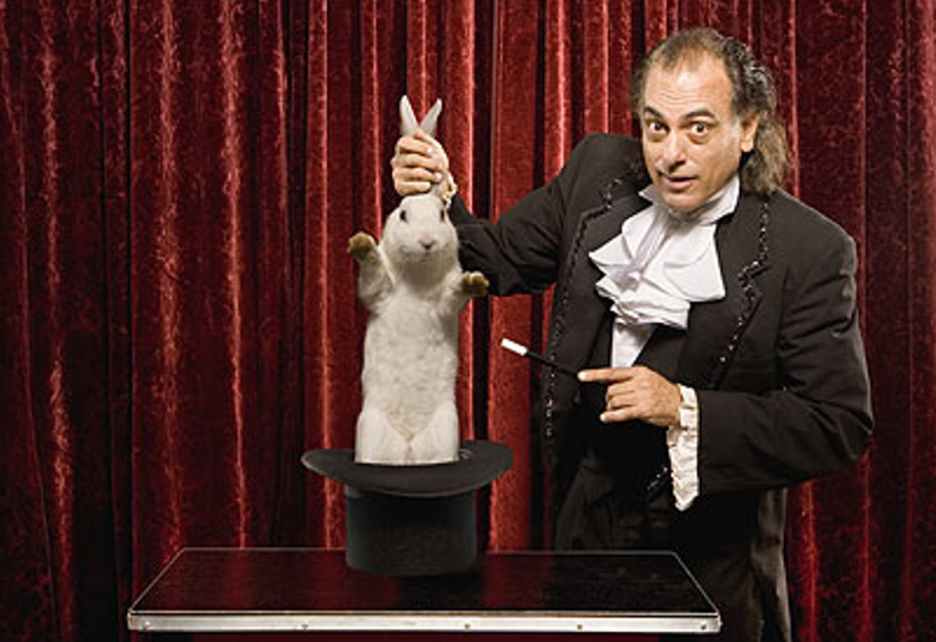}
\caption{Science works like magic, nay better than magic, even better than science fiction...}
\label{fig:magic}
\end{figure}

\noindent
Physics indeed works like magic. We can from our today’s physics produce apparatuses which would bluff any back-in-time physicist, activate devices at distance (and soon at the motion of an eyelid, and even at some point from a mere thought), and other feats (See Figure~\ref{fig:magic}).

In fact, I would say it may work better than magic as we not only manipulate objects, mass produce them, improve on them, but crucially we understand their mode of operation. We even do often better than science fiction: Science fiction imagines things while science not only conceive them, but it explains and produce them. Quite often, objects manufactured by science couldn’t have been thought by science fiction writers. 

In the same vein, not to leave anyone unaccounted for, chemists do better than alchemists: They produces wondrous substances with properties ``\`a la carte''. They are the magicians of the molecular world.

\section{Physics as the Grammar of Nature}
\label{sec:grammar}
\begin{figure}[h]
\includegraphics[width=\textwidth]{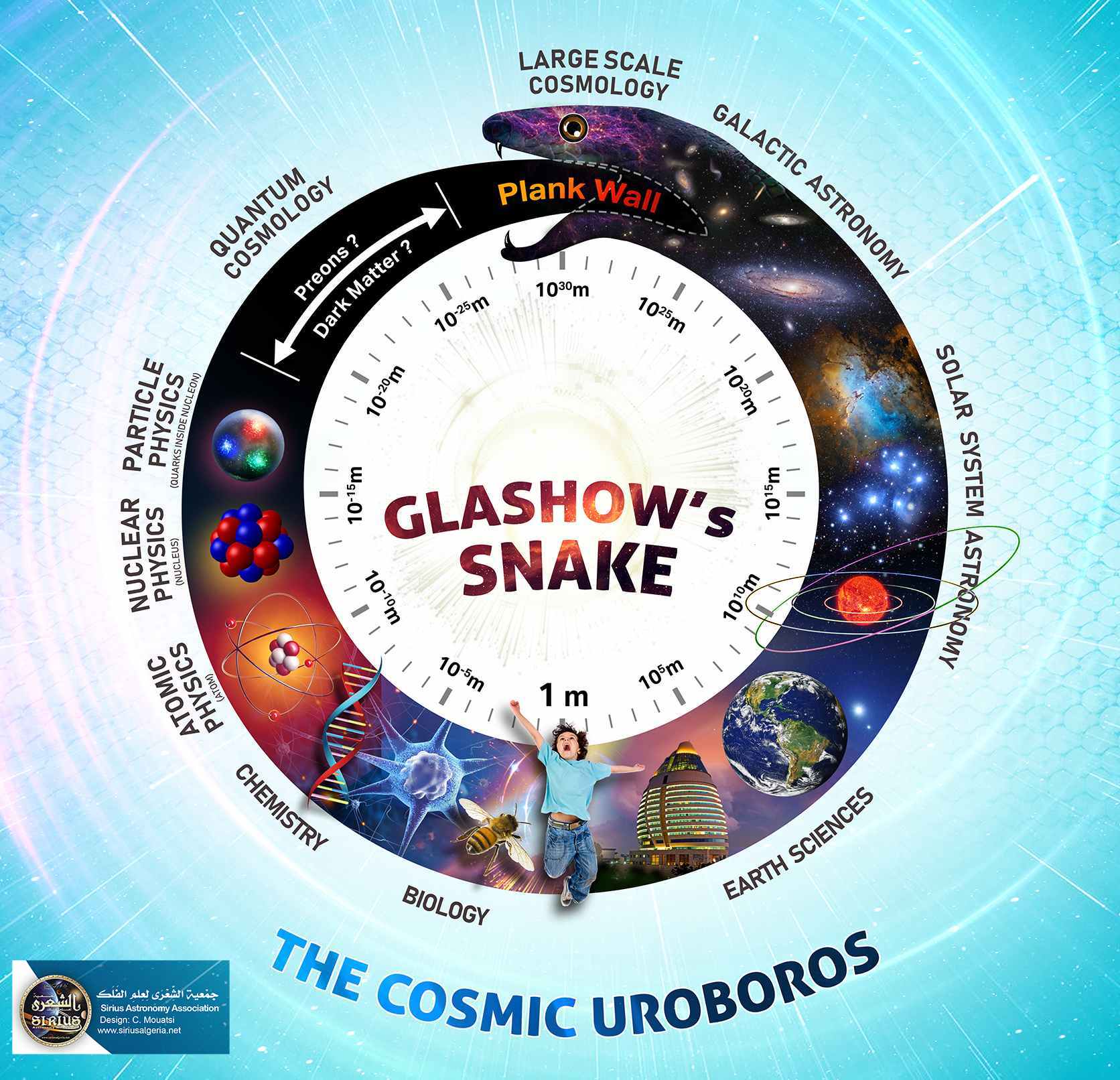}
\caption{The Cosmic Uroboros showing the two infinities and their merging}
\label{fig:snake}
\end{figure}

\noindent
Let us ask the rhetorical and somewhat provocative question: What is the most basic science of all? It is certainly the science which deals with the intimate nature of material things at all scales! This is precisely what physics is about and on this I rest my case that it is indeed the most fundamental science and I may even add, by definition! It provides at each scale a description of what is going on, makes quantitative predictions and predicts correctly outcomes coming out of the experimentally tested domain.
Its domain of action covers all the space and time scales albeit with different degrees of completion for some. We can illustrate this in two ways, one through the Glashow’s snake or the cosmic Oroboros which not only shows the vast domain of relevancy of physics from the infinitely small (particle physics down to the Planck scale, at least in principle) up to the extra galactic realm, with the additional benefit of portraying the new deal where cosmology “merges” with sub-nuclear physics, achieving the coming together of the two infinities (Figure~\ref{fig:snake}), 

\begin{figure}[h]
\includegraphics[width=\textwidth]{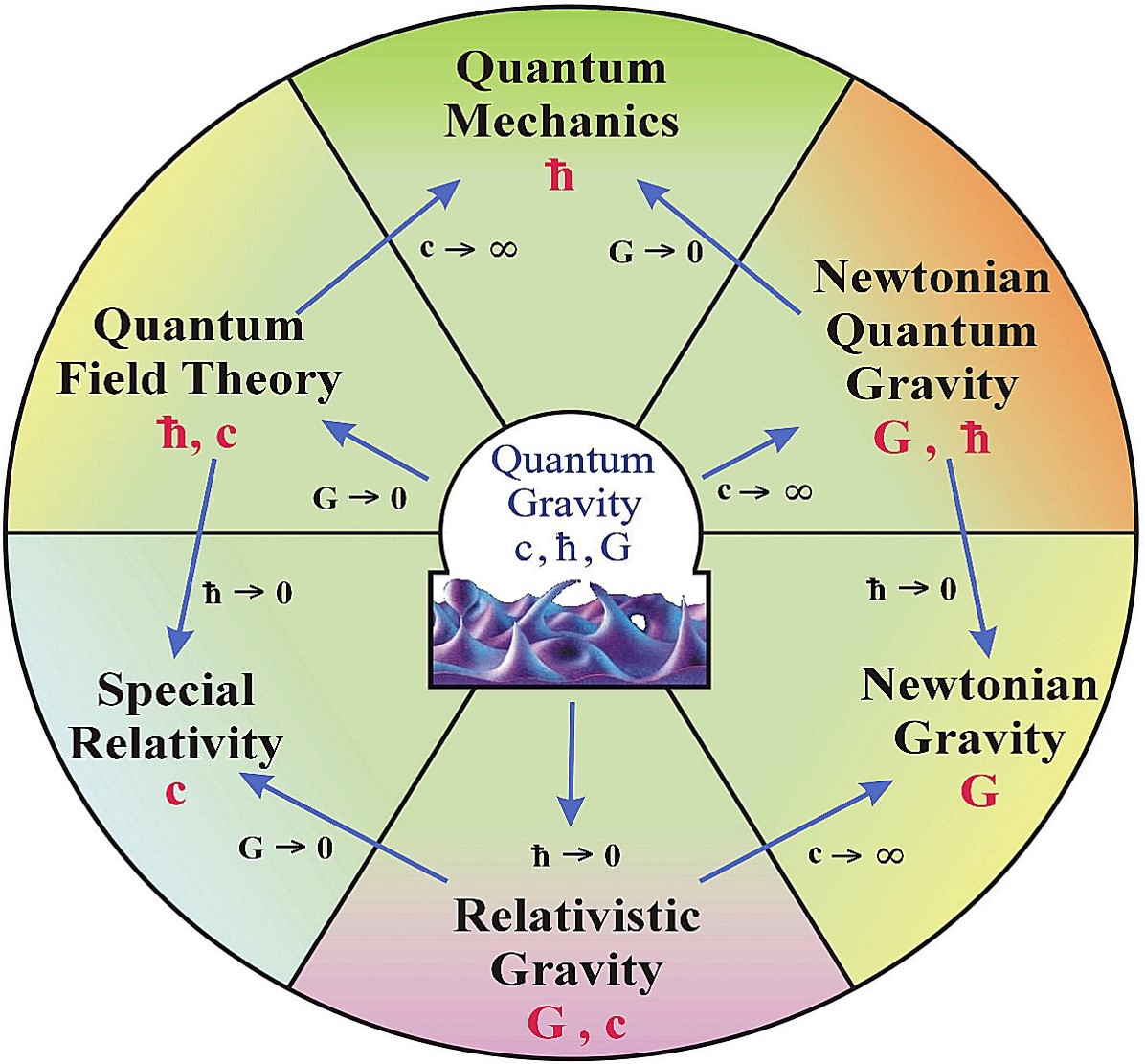}
\caption{The art of taking the limits}
\label{fig:wheel}
\end{figure}

Physics is also this fundamental science used by a multitude of disciplines. Sometime we hear a rather similar claim about Mathematics. Well, keeping in the realm of analogies, if Mathematics is the language of Nature, Physics is its syntax and grammar, its underpinnings. Mathematics is not unlike a toolbox, but if using the right tool is crucial for the car mechanic, a toolbox without a motor to apply it to is useless. Physics is in fact so basic that we are not enough conscious of its fundamental character! Let us list some of those disciplines which use it in a fundamental way: 

We can also see the efficiency of physics through this wheel of approximations in the diagram below. See Figure~\ref{fig:wheel} By assigning various values to the fundamental constants, we can blithely hopping through the various domains of validity of covering all of the physical world at various regimes. It is the ultimate game of taking the limits that theoretical physicists have brought to a high degree of perfection. It is also the science which strives to synthetize all the knowledge within it in a grand unified explanation scheme, quite a lofty goal for any science. It is in fact the only science which pushes the fullest unification attempt of its domain of investigation as one of its goals (The theorists at least)… even if this objective has been looming farther and farther during the past several decades.

- The Earth Sciences, from geophysics to atmospheric physics, to oceanography, to plate tectonics...

- Life Sciences, from biophysics and medical physics of course, to plant physiology, cardiology… It is used for example in cytology to model the movement of the cell through the beating of cilia, for following the motion of substances across the cytoplasmic membrane…

- The Sciences of the Universe as it is obvious enough: Is not the Universe the ultimate laboratory of matter under diverse physical conditions including the most extreme ones? 
We may add to the list the various branches of engineering, which behind their facade of utilitarianism, are in fact practicing the art of optimizing cost (with the exigency of availability of ingredients, ergonomics of the product, public taste…) using physics! It may indeed be described as “opportunistic physics” or more politely “constrained physics”.

Thus, much like we cannot do without mathematics to analyze and interpret data, it is too not possible to dispense ourselves of physics to do science, almost any science, even if we wanted to. Physics is indeed for the other sciences what the air we breathe is for life.

\section{Some of the Physics’ success stories}
\label{sec:success}
\noindent
Every scientific discipline which ever appeared has been brought to fruition and has in the process developed a unique practice and understanding that no one not trained within it can compete with. Physicists can rightly be proud of having achieved great successes. I would like to point out to three such broad domains although I won’t be able to expand on any of them here. I will choose them at very different scales:

- Subatomic scale: We have been able to map out the particle content of the sub-nuclear world and its modus operandi (interactions, conservation laws…) through mainly the Standard Model.

- Atomic scale: Starting from Schrodinger equation and the orbital configuration business, i.e. the whole chemistry, we can confidently go about predicting properties of any atomic and molecular structure one can think of, at least to some good degree of accuracy. Explaining the stellar atmosphere spectra is also a tribute to the mastery we have reached, and by the way, a success story of statistical mechanics not often acknowledged.

- Stellar world: the physics of the stars and their evolution from their birth to their demise, has matured to the point where we can explain the Hertzsprung - Russell diagram: location, density and timeline of the various type of stars represented in the diagram. Yet many surprises still await us with stars of weird constitution not yet discovered.

\section{Imperialistic … but generous}
\label{sec:imperialist}
\begin{figure}[h]
\includegraphics[width=\textwidth]{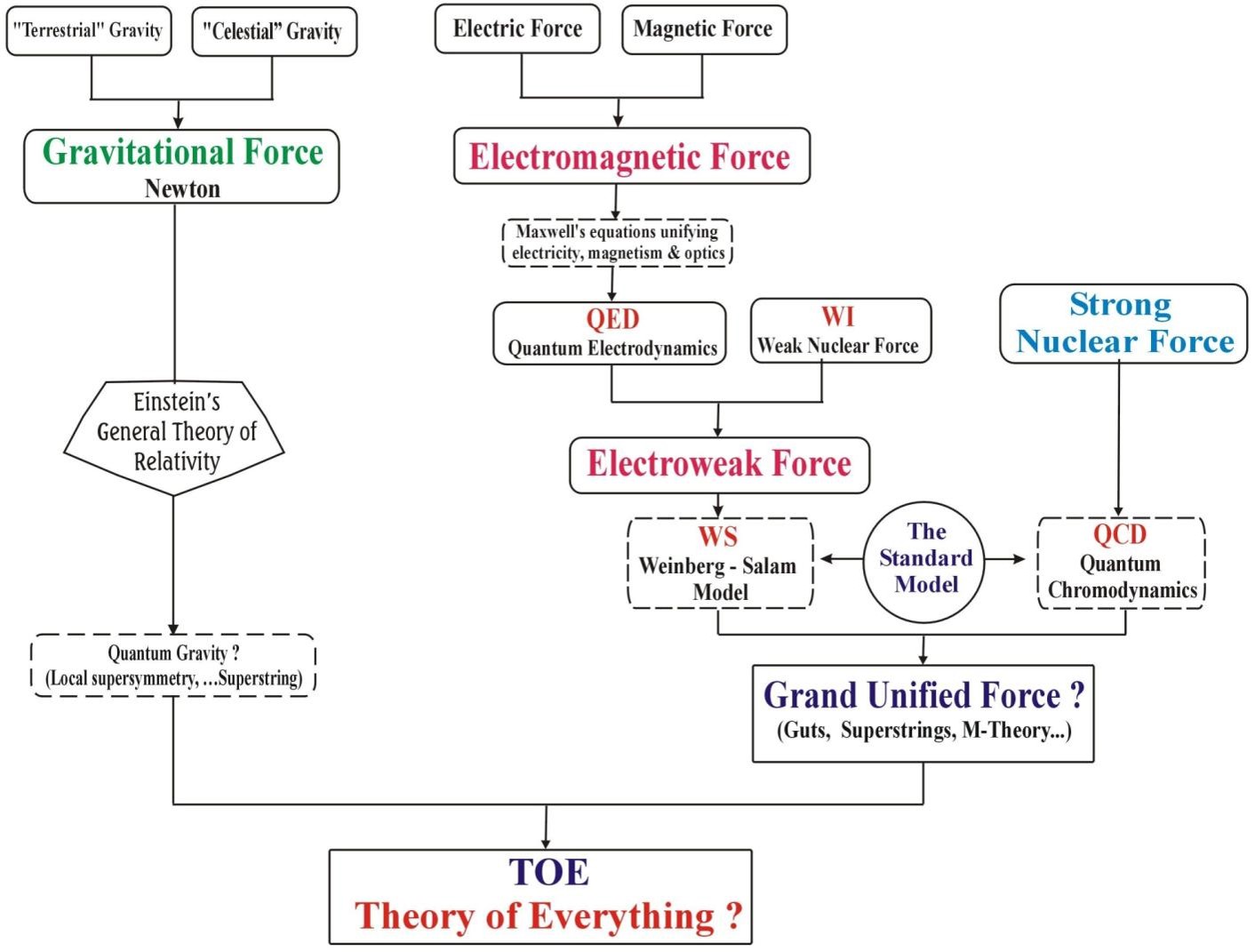}
\caption{Towards a Theory of Everything ?}
~\label{fig:everything}
\end{figure}

\noindent
Physics is characterized by a very specific spirit, that of magnanimity and selflessness. In fact, the imperialistic tendency often attributed to it is really an unsubstantiated feeling. Physicists are a race of pioneers who clear up constantly new territories up to the borders of the unknown, then leave to other disciplines the task to exploit them. While it does occupy basically all the domains of the material world as we saw, at all scales, from the microscopic, to the macroscopic, to the cosmic realm, yet the physicist's curiosity is fully disinterested. There is indeed no "occupation de terrain"; it generates cutting-edge knowledge but leaves to others the task of making it flourish and find applications, thus electronics, nuclear energy, laser technology\footnote{\label{laser}It spurred the development of whole branches of mathematics like the theory of distribution (Quantum mechanics), the calculus of variations (Lagrangian mechanics and QFT…), Riemannian and non-Riemannian geometries (General Relativity…) … and I would dare add the Web, an invention at CERN, a spin-off of accelerator physics meant to improve communications and the exchange of data across collaborating centers all over the world.} (Figure~\ref{fig:everything}).

Claims have been made that chemistry has been annexed to it becoming a subdomain with a special vested interest\footnote{\label{Chemistry annexed}Yet we understand that this situation arises from the natural way of classification of sciences which goes back to the meanders of the history of the “physical sciences”. If one is to feel better about it, call the chemist an applied molecular physicist, although I doubt that will improve the situation… In any case, the overlapping is there and the rational for it is that physicists and chemists working in this common domain have different agendas.}. Actually it is true in principle that physics has been tasked to unravel the working of the whole material world, and so the other material branches have been made de facto subsidiary to it, but that's from the hazards of the history of sciences. Yet physics is in compensation generous. It hands over any new branch it opens to technologists to fructify it.

\section{Physics in a Nutshell}
\label{sec:nutshell}
\noindent
The magic of physics lays in the utter simplicity it goes about explaining the world. The fundamentals laws are a concentrate of sobriety to the point of simpleness, of coherence, and (mathematical) eloquence, in addition to elegance\footnote{\label{No Truth}Some people may argue that laws, and thus a theory built around its equations merely embodies a given set of observations and thus has no “truth” in it. We can over go further in relativizing the meaning of truth to turn into a “human truth” like in Bohr’s words: “It is wrong to think that the task of physics is to find out how nature is. Physics concerns what we can say about Nature.” In other words, physics is a model of our perception of the world, but not the world itself as much as the map is not the territory. Actually, there are different ways from an epistemological point of view of understanding the core principles of physics, as there are many paths to spirituality…}. 
Look at mechanics which is set to explain most of the large scale behavior of matter. What used to be the biggest stumbling block to human thinking, namely the nature of motion, now embodied in the first law of Newton, ultimately states according to Galileo that motion is like ``nothing''. As for the remaining stuff embodied in the other Newton’s laws\footnote{\label{Newton's law}The second Newton law states that an applied force to a body provides an acceleration impeded only by the “inertia” of its mass, while the third one is just the push–pull like rule that to every action there is a reaction equal and opposite.}  which can explain the trajectories of intercontinental missiles, the stability of bridges, the running of plasma engines, the orbits of planets, the motions of stars in clusters, it just operationally states: 

-  Search for a deviation from the inertial motion (the uniform linear motion, or “The motion is like nothing “), which would embody the presence of an (external) force!

- Plug in the expression of the force into Newton's second law, and you get the Equations of Motion (EoM)… and then go solve them!

If Newton laws’ simplicity didn’t impress you, consider electromagnetism where every electrical, magnetic or radiation phenomenon you have ever heard of can be deduced from the four Maxwell differential equations (See the didactic challenges below). The whole of geometric optics is but the eikonal approximation applied to those equations. Or alternatively, look at General Relativity (GR), especially if derived from a least action principle; it is the epitome of aesthetic beauty and simplicity\footnote{\label{QM}Quantum mechanics could also be formulated in a terse axiomatic mode, but the non-traditional aspects of the theory call for some mathematical culture so as to fully comprehend its “abstruseness” and translate it into physically “operational” terms.}…

Of course in practice, each of those laws have to be seriously unpacked before been made full use of. For example Einstein’s equations of GR is a set of highly non-linear set of coupled tensor differential equations… which can’t be solved exactly except in the simplest cases (Like Black Holes…). 

\begin{figure}
\includegraphics[width=\textwidth]{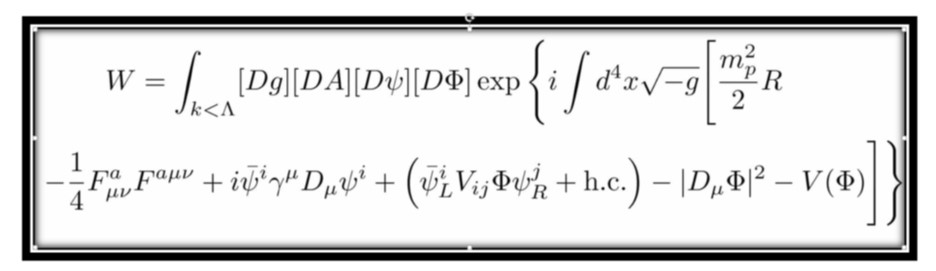}
\caption{The equation of the "World" in a path integral formulation. There is the space-time structure, the material content and all the fundamental interactions}
~\label{fig:pathequation}
\end{figure}

In order to convey further the idea that simplicity doesn’t mean ease of manipulation, here’s the ultimate theory of everything (That we know of today…) written in a compact path integral formulation (Figure~\ref{fig:pathequation}). May we add that a whole (theoretical) life of unpacking may be needed to explore the various terms. This equation indeed contains in a compact way all the known microscopic physics. Yet it by no mean a Theory of Everything. Thus it doesn’t explain how the various material entities represented in the various terms above emerge or are related to each other. It is just an equation containing all the basic interactions in a unified way, formulated as a gauge invariant theory and written in the path integral formalism.

Thus physics ultimately embodies the power of simplicity and parsimony applied to slick fundamental objects. At the end of the road, there is an incredible accuracy of its predictions when confronted to experimental data. Explaining a very large set of phenomena with a minimum set of principles is very much like the artistic expression of an esthete pushed to the extreme. Theoretical physics in some strong sense is an art, and Newton, Maxwell, Einstein, Dirac are their emblematic heroes. 

\begin{figure}[h]
\includegraphics[width=\textwidth]{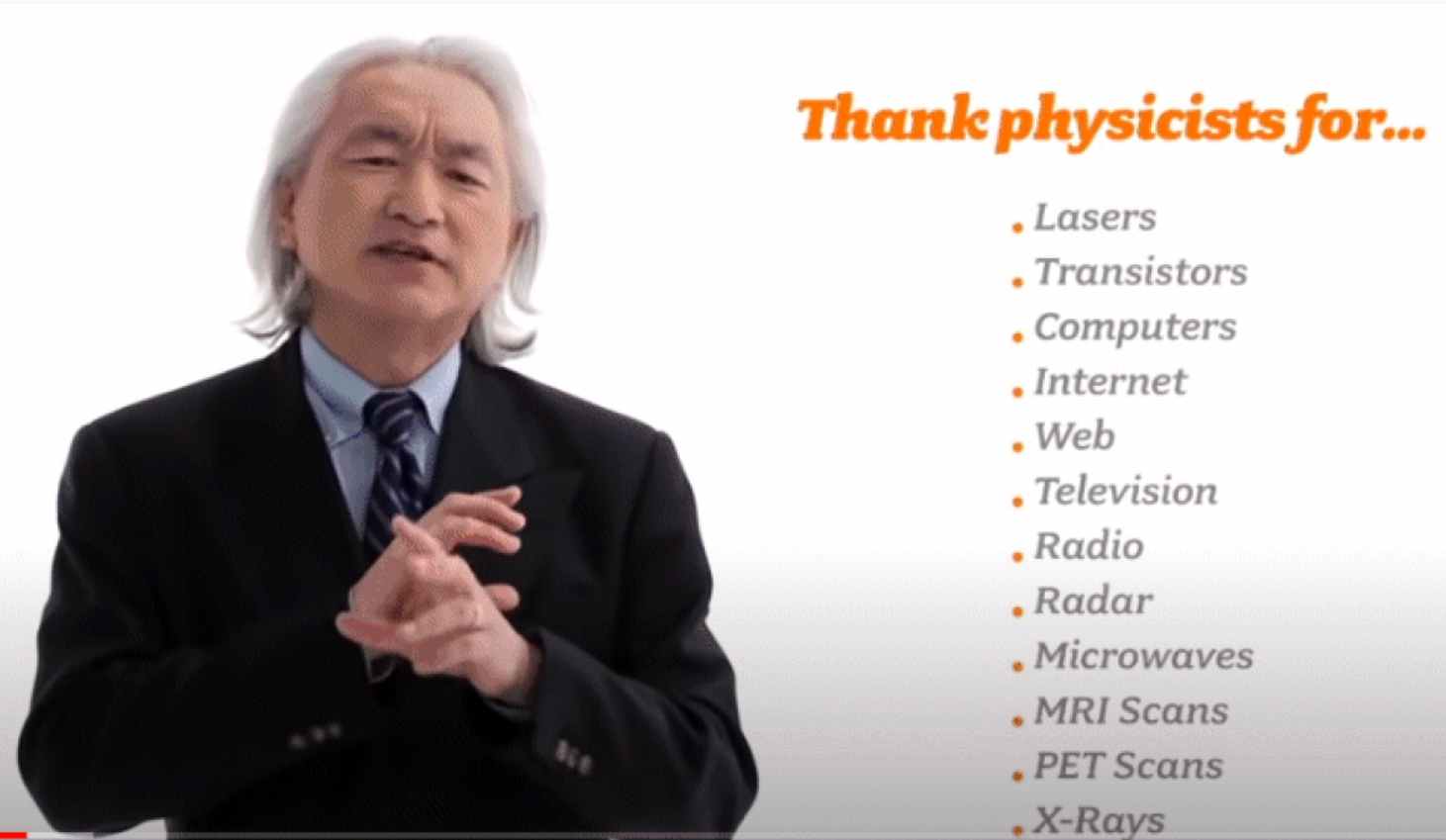}
\caption{Slide taken from a lecture by theoretical physicist Michio Kaku from City College of New York}
~\label{fig:Kaku}
\end{figure}

Furthermore, the basic laws become even simpler and the description easier when going to smaller scale (thus to higher energy), or back in time. Look as an example at the behavior of the quarks and the asymptotic freedom they gain at high energy originating from the mathematical structure of QCD, or the structure of matter when getting closer to the Big Bang. Even the Black Hole which was thought for a long time to be the simplest object in the Universe as it is characterized with only two numbers namely it’s mass and spin speed
In addition to those feats, physics has much to claim from a utilitarian point of view, but this is so well known that there should be no need to elaborate further here (Figure~\ref{fig:Kaku}).

\section{The Didactic Challenge and about Closing the epistemic gap }
\label{sec:didactic}
\noindent
Physics for what it's worth is all in its conceptual part. Without this cognitive dimension which enables one to truly understand the material world around us, it would be an unpretentious phenomenological discipline that is just a degree above the usual explanations in non-fundamental disciplines. It is that cognitive part which can fire up the imagination of aspirant physicists and kindle genuine vocations. Take away this conceptual part and you draw closer to be soldiers of fortune of physics than true physicists. It is all the difference between mercenaries and freedom fighters. Indeed, if all one’s skills is to apply equations and crunch numbers, one is not doing physics but is just at best a physics consumer. This is precisely the problem with all those bright young minds who get their baccalaureate (Bac) with flying scores in physics… and then go to medicine. They haven’t mastered physics for the least but have just worked out all the possible combinations of problems that may come to the exam according to the chapters covered\footnote{\label{Pandemic}Yet its description might be a daunting task if not an impossible one. First because lurks at its center a singularity that we don’t know how to properly describe physically. Then because the devil of thermodynamics has stepped in and now we are not sure of what is ambushed just behind the Schwarzschild horizon. A “firewall” theory was proposed in 2012, namely a hypothetical phenomenon where an observer falling into a black hole encounters high-energy quanta close or at the event horizon and get fully roasted. Never mind the subsequent spaghettification, if he ever reaches that stage..}.
 Luckily books have appeared, although not much known or used in our countries in Africa, which appeal to the conceptual part of physics
 ~\cite{epstein2002thinking}
~\cite{hewitt2014conceptual}
~\cite{adelson2007flying}.
The famous Feynman's lectures in physics ~\cite{feynman2011feynman} have been and still are a goldmine for those who wish to go into the inner working of physics and look for physical insight. 
Some have adopted the precise approach I am advocating for here like Hassani's one~\cite{hassani2010atoms} who states in his preface:
 
 \textit{"This book goes against the philosophy that the only physics worth learning is that which can be applied to the making of a cell phone, high definition TV, computer, SUV, or movie filled with graphics trickery. It does not glorify the pragmatic aspect of physics. Rather, it talks about some of the great ideas that have defined our era; about the inner workings of that magnificent enterprise of our race, science; about the power of theoretical prediction and the delight of experimental confirmation..."}
 
 Nothing better to show the deficiency in physics understanding than to have would-be physicists (even bona fide ones) go through a series of didactic challenges where the skill is not to regurgitate a theory or an equation, but to explain. Here are few of them:
 
- What is the most fundamental science of all? Luckily, I have already answered that one.

- Why is there only two electrons in the first layer of atoms?

- By which interaction is the Sun shining? 

-  Describe braking thermodynamically wise.

- Explain all magnetic phenomena in one sentence ... and all e.m. wave production in another one.

- Why doesn’t an inclining spinning top fall over?
  
 - Explain the operation of spinning on oneself, that is of changing your direction of motion when standing on Earth’s surface.
 
\begin{figure}
\centering
\includegraphics[width=8cm]{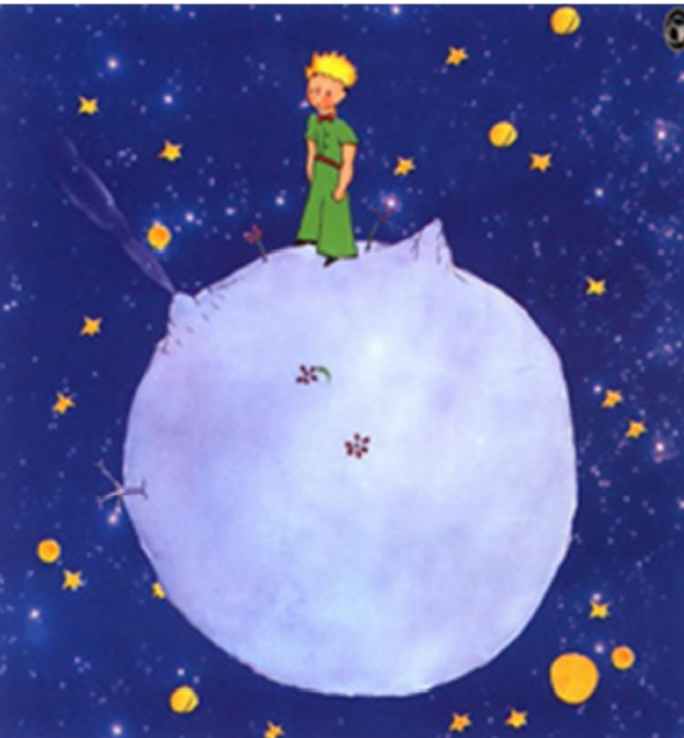}
\caption{Explaining the operation of spinning on oneself: Here's some help from the story of ``le Petit Prince'' of Saint-Exupery}
~\label{fig:prince}
\end{figure}

 For this later one, a Newtonian physicist’s explanation would go like that: Go about rotating the Earth in the opposite direction of the one you want to spin, and in reaction, the Earth will make you rotate in the right direction! The Earth is huge so its rotation will be imperceptible while you, tiny creature, will rotate the way you intended. I have no place to explain in more detail here why, but I will display a candid version from an illustration for “Le Petit Prince” of Saint-Exupery which should be self-explanatory (Figure~\ref{fig:prince}).
 
 Some answers to those above questions may look paradoxical if you use your “common sense” logic, which is precisely what a physicist shouldn’t go about.

\section{Grand connections and inner workings}
\label{sec:connections}
\noindent
Let us go back to our original questioning: Why don't people understand physics? The answer I argued about was that because physics has never been made attractive, and in the first place by physicists themselves. It is as if physicists all too often in their own hearts and minds believe that indeed their stuff is boring. So they take solace in going into problem solving and exerting their guts in developing the best bag of tricks to faster find solutions to some normative problems. They are in some way applying subconsciously what Feynman stated: “Nobody understands quantum mechanics”\footnote{\label{Beauty}This statement of Feynman was pronounced in connection with the puzzle of the two-slit experiment. In addition to be a candid acknowledgement of the weirdness of quantum mechanics in comparison to all the other branches of physics, it is more an attitude that when theorizing on a topic, namely that we should submit to rock solid facts no matter how strange they might look, and it is certainly not an attitude of powerlessness if not despair as displayed by some meek physicists.}, but this time to the whole of physics and in a defeatist mood.
Actually if physicists want to save their discipline or to be at least true to its spirit, and not to behave as disembodied soulless physicists, they should become essentialists. Namely go: back to basics! Every discussion on a physics related issue should always point back to its fundamental inner working. Talking about artificial satellites that’s Newton centripetal force in action, or in less abstract term, that’s free fall around a spherical Earth. But don’t obfuscate either: the blueness of the sky should not be explained away by merely uttering the magic answer - “Raleight’s scattering”- or by bringing to bear some quantum theory of light propagation in a medium (Unless you have the proper audience for that), but as resulting from some  easily understood soft semi-classical exposition.
\begin{wrapfigure}{r}{0.45\textwidth}
    \centering
    \includegraphics[width=0.45\textwidth]{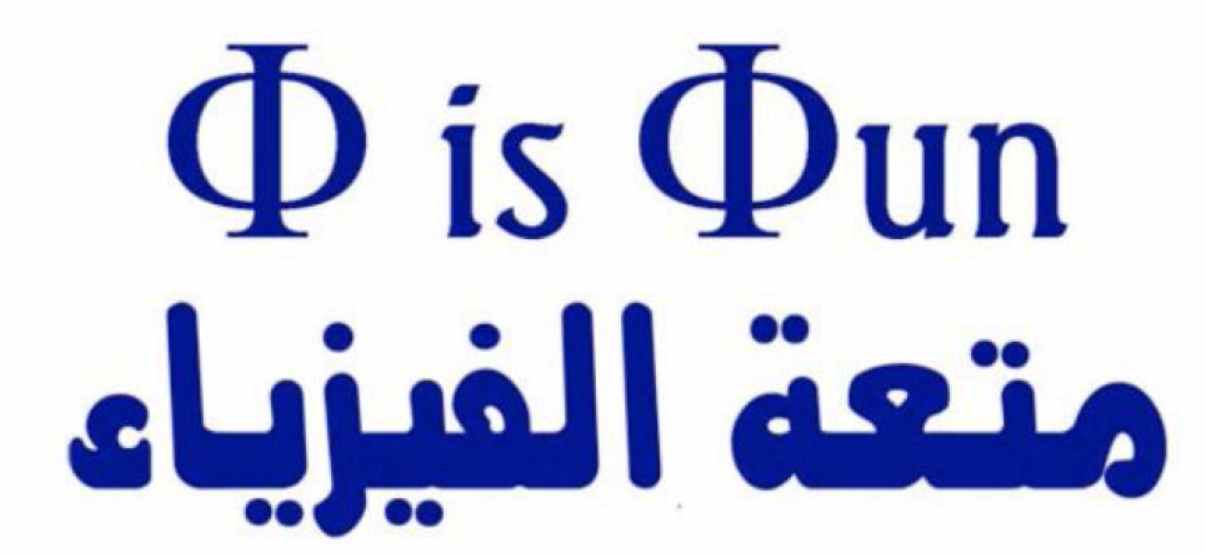}
\end{wrapfigure}
Like maternal love, you don’t necessary increase the feeling of filial love by rehashing a series of moral rules and duties that the society has set up, but by appealing to the inner sentiments towards your Mom, from which the rules are but a pale copy, not unlike the shadows in Plato's allegory of the cave. 

Or like the beauty of a mathematical theory, it is not the range of its applications whatever extended it might be which counts, nor the elegance of some of the demonstrations, that’s all fioritura; but rather in the purity of its minimalist set of axioms and the richness of developments and connections it allows.

If physicists invoke the grand picture, the mighty connections, the coherence and the simplicity, then utilitarian arguments become quite secondary. That’s at least the proper spirit of physics, and that calls for an extra dose of love for one’s own discipline. In this process of making a case for physics, he might convoke some high ends of physics that he might not master (One needs not know about the quark model or QCD in order to state that nucleons are made of sub-entities called quarks...), while de-emphasizing the utilitarian and problem solving aspects of his discipline.

\newpage

\bibliographystyle{elsarticle-num}
\bibliography{acp2021-Mimouni-Jamal} 
\end{document}